# A Riccati-type solution of 3D Euler equations for incompressible flow

**Sergey V. Ershkov**

Nizhny Novgorod State Technical University n.a. R.E. Alekseev,

24 Minina st., Nizhny Novgorod 603155, Russia

e-mail: sergej-ershkov@yandex.ru

**Roman V. Shamin**

Peoples' Friendship University of Russia,

6, Miklukho-Maklay Street, Moscow 117198, Russia,

e-mail: roman@shamin.ru

In fluid mechanics, a lot of authors have been reporting analytical solutions of Euler and Navier-Stokes equations. But there is an essential deficiency of non-stationary solutions indeed.

In our presentation, we explore the case of non-stationary flows of the Euler equations for incompressible fluids, which should conserve the *Bernoulli*-function to be invariant for the aforementioned system.

We use previously suggested ansatz for solving of the system of Navier-Stokes equations (which is proved to have *the analytical* way to present its solution in case of conserving the *Bernoulli*-function to be invariant for such the type of the flows). Conditions for the existence of exact solution of the aforementioned type for the Euler equations are obtained. The restrictions at choosing of the form of the 3D nonstationary solution for the given constant *Bernoulli*-function $B$ are considered. We should especially note that pressure field should be calculated from the given constant *Bernoulli*-function $B$, if all the components of velocity field are obtained.

**Keywords:** Euler equations, *Bernoulli*-function, non-stationary solutions.

1. **Introduction, the Euler system of equations.**

The Euler system of equations for incompressible flow of inviscid fluid is known to be one of the old famous problems in classical fluid mechanics, besides we should especially note that a lot of great scientists have been trying to solve such a problem during last 265 years.

In accordance with [1-3], the Euler system of equations for incompressible flow of inviscid fluid should be presented in the Cartesian coordinates as below (under the proper initial or boundary conditions):

$$\nabla \cdot \vec{u} = 0, \qquad (1)$$

$$\frac{\partial \vec{u}}{\partial t} + (\vec{u} \cdot \nabla)\vec{u} = -\frac{\nabla p}{\rho} + \vec{F}, \qquad (2)$$

- where $u$ is the flow velocity, a vector field; $\rho$ is the fluid density, $p$ is the pressure, $F$ represents external force (*per unit of mass in a volume*) acting on the fluid; besides, we assume that external force $F$ above has a potential $\phi$ represented by $F = -\nabla \phi$.

2. **The originating system of PDE for Euler Eqs.**

Using the identity $(u \cdot \nabla)u = (1/2)\nabla(u^2) - u \times (\nabla \times u)$, we could present equations (1)-(2) in case of incompressible flow $u = \{u_1, u_2, u_3\}$ as below [4-5]:

$$\nabla \cdot \vec{u} = 0,$$

$$\frac{\partial \vec{u}}{\partial t} = \vec{u} \times \vec{w} - \nabla B \qquad (3)$$

- where *Bernoulli*-function $B$ is given by the appropriate expression below:

$$B = \frac{1}{2}(\vec{u}^2) + p + \phi,$$

- also we denote in (3) *the curl field* **w** = ∇×**u**, a pseudovector field (*time-dependent*) [6]:

$$\{w_1, w_2, w_3\} \equiv \left\{ \left( \frac{\partial u_3}{\partial y} - \frac{\partial u_2}{\partial z} \right), \left( \frac{\partial u_1}{\partial z} - \frac{\partial u_3}{\partial x} \right), \left( \frac{\partial u_2}{\partial x} - \frac{\partial u_1}{\partial y} \right) \right\} \quad (4)$$

Let us search for solutions {**u**, *p*} of the system of equations (3) which should conserve the *Bernoulli*-function to be the invariant for the aforementioned system [7]:

$$B = \frac{1}{2}(\vec{u}^2) + p + \phi = const \quad (5)$$

Basically, the aforementioned assumption (5) is associated in fluid mechanics text books with the case of incompressible flow under the two additional demands as below:

1. Steady inviscid flow (the solution does not depend on time);
2. Irrotational flow (the fluid particles don't spin, i.e., the curl of velocity is zero).

We consider here the case of non-stationary solution of inviscid rotational flow (i.e., the case with non-zero curl of velocity).

### 3. **Presentation of the time-dependent part of solution.**

The second of the equations of system (3) with the additional demand (5) is proved to have *the analytical* way to present its solution [8-10] in a form below (in regard to the time-parameter *t*):

$$u_1 = -\gamma \cdot \left( \frac{2a}{1+(a^2+b^2)} \right), \quad u_2 = -\gamma \cdot \left( \frac{2b}{1+(a^2+b^2)} \right),$$

$$u_3 = \gamma \cdot \left( \frac{1-(a^2+b^2)}{1+(a^2+b^2)} \right) = \gamma \cdot \left( \frac{2}{1+(a^2+b^2)} - 1 \right), \quad (6)$$

- where γ (*x*, *y*, *z*) is some arbitrary function, given by the initial conditions; the real-valued coefficients *a*(*t*), *b*(*t*) are solutions of the mutual system of 2 *Riccati* ordinary differential equations (let us consider variables {*x,y,z*} as variable parameters below):

$$\begin{cases} a' = \dfrac{w_2}{2} \cdot a^2 - (w_1 \cdot b) \cdot a - \dfrac{w_2}{2}(b^2 - 1) + w_3 \cdot b, \\ \\ b' = -\dfrac{w_1}{2} \cdot b^2 + (w_2 \cdot a) \cdot b + \dfrac{w_1}{2} \cdot (a^2 - 1) - w_3 \cdot a. \end{cases} \quad (7)$$

It is clear that for the correct solving of system (3) with additional demand (5) we should obtain from Eqs. (4), (6) the three additional equations below (which should determine the spatial part of *the curl field **w***):

$$\{w_1, w_2, w_3\} = \left\{ \frac{\partial\left(\gamma \cdot \left(\frac{2}{1+(a^2+b^2)} - 1\right)\right)}{\partial y} + \frac{\partial\left(\gamma \cdot \left(\frac{2b}{1+(a^2+b^2)}\right)\right)}{\partial z}, \; -\frac{\partial\left(\gamma \cdot \left(\frac{2a}{1+(a^2+b^2)}\right)\right)}{\partial z} - \frac{\partial\left(\gamma \cdot \left(\frac{2}{1+(a^2+b^2)} - 1\right)\right)}{\partial x}, \right.$$

$$\left. -\frac{\partial\left(\gamma \cdot \left(\frac{2b}{1+(a^2+b^2)}\right)\right)}{\partial x} + \frac{\partial\left(\gamma \cdot \left(\frac{2a}{1+(a^2+b^2)}\right)\right)}{\partial y} \right\} \quad (8)$$

Besides, using the continuity equation (1), we should obtain from (6):

$$-\frac{\partial\left(\gamma \cdot \left(\frac{2a}{1+(a^2+b^2)}\right)\right)}{\partial x} - \frac{\partial\left(\gamma \cdot \left(\frac{2b}{1+(a^2+b^2)}\right)\right)}{\partial y} + \frac{\partial\left(\gamma \cdot \left(\frac{2}{1+(a^2+b^2)} - 1\right)\right)}{\partial z} = 0, \quad (9)$$

For the sake of simplicity, let us search for solutions (6) according to the ansatz pointed below:

$$\frac{\partial\left(\gamma\cdot\left(\frac{2}{1+(a^2+b^2)}-1\right)\right)}{\partial z}=0, \quad \Rightarrow \quad \gamma\cdot\left(\frac{2}{1+(a^2+b^2)}-1\right)=\sigma(x,y,t), \quad \Rightarrow$$

$$(a^2+b^2)=\left(\frac{1-\frac{\sigma(x,y,t)}{\gamma(x,y,z)}}{1+\frac{\sigma(x,y,t)}{\gamma(x,y,z)}}\right), \quad \Rightarrow \quad \frac{1}{1+(a^2+b^2)}=\frac{1}{1+\left(\frac{1-\frac{\sigma(x,y,t)}{\gamma(x,y,z)}}{1+\frac{\sigma(x,y,t)}{\gamma(x,y,z)}}\right)}=\frac{\left(1+\frac{\sigma(x,y,t)}{\gamma(x,y,z)}\right)}{2},$$

- besides, let us additionally assume

$$\frac{\partial\left(\gamma\cdot\left(\frac{2b}{1+(a^2+b^2)}\right)\right)}{\partial y}=0, \quad \Rightarrow \quad \gamma\cdot\left(\frac{2b}{1+(a^2+b^2)}\right)=\lambda(x,z,t), \quad \Rightarrow \quad b=\frac{\lambda(x,z,t)}{(\gamma(x,y,z)+\sigma(x,y,t))},$$

$$\Rightarrow \quad a=\frac{\sqrt{(\gamma^2(x,y,z)-\sigma^2(x,y,t)-\lambda^2(x,z,t))}}{\gamma(x,y,z)+\sigma(x,y,t)}, \qquad \gamma^2(x,y,z)-\sigma^2(x,y,t)-\lambda^2(x,z,t)\geq 0 \qquad (10)$$

- so, it means from the analysis of Eqs. (9)-(10) that ($A = A(t)$, $\gamma^2 > \sigma^2$):

$$\frac{\partial\left(\gamma\cdot\left(2a\frac{\left(1+\frac{\sigma(x,y,t)}{\gamma(x,y,z)}\right)}{2}\right)\right)}{\partial x}=0, \quad \Rightarrow \quad \frac{\partial\left(\sqrt{(\gamma^2(x,y,z)-\sigma^2(x,y,t)-\lambda^2(x,z,t))}\right)}{\partial x}=0,$$

$$\Rightarrow \quad \gamma(x,y,z)\equiv\gamma(y,z), \quad \sigma(x,y,t)\equiv\sin(A(t)\cdot x)\cdot\sigma(t), \quad \lambda(x,z,t)\equiv\cos(A(t)\cdot x)\cdot\sigma(t), \qquad (11)$$

- or, for example

$$\Rightarrow \quad \gamma(x,y,z) \equiv \gamma(y,z), \quad \sigma(x,y,t) \equiv \sigma(y,t), \quad \lambda(x,z,t) \equiv \lambda(z,t), \qquad (12)$$

- but in the last variant of exact solution (which should satisfy to the continuity equation (9)) variable *x* is eliminated from the presentation of the solution, so such the solution is simply reduced to the 2D non-stationary solution in case of (12).

Thus, let us search for the exact solution in a form (11); in this case, expressions for the components of the curl field (8) should be presented as below:

$$\{w_1, w_2, w_3\} = \left\{ \frac{\partial(\sigma(x,y,t))}{\partial y} + \frac{\partial(\lambda(x,z,t))}{\partial z}, \; -\frac{\partial\left(\sqrt{\gamma^2(x,y,z)-\sigma^2(x,y,t)-\lambda^2(x,z,t)}\right)}{\partial z} - \frac{\partial(\sigma(x,y,t))}{\partial x}, \right.$$

$$\left. -\frac{\partial(\lambda(x,z,t))}{\partial x} + \frac{\partial\left(\sqrt{\gamma^2(x,y,z)-\sigma^2(x,y,t)-\lambda^2(x,z,t)}\right)}{\partial y} \right\}$$

- which, taking into account expressions (10)-(11), should be transformed to the form below ($\gamma^2 > \sigma^2$):

$$\gamma(x,y,z) \equiv \gamma(y,z), \quad \sigma \equiv \sin(A(t)\cdot x)\cdot\sigma(t), \quad \lambda \equiv \cos(A(t)\cdot x)\cdot\sigma(t) \quad \Rightarrow$$

$$\{w_1, w_2, w_3\} = \left\{ 0, \; -\frac{\partial\left(\sqrt{\gamma^2(y,z)-\sigma^2(t)}\right)}{\partial z} - A(t)\cdot\cos(A(t)\cdot x)\cdot\sigma(t), \right.$$

$$\left. A(t)\cdot\sin(A(t)\cdot x)\cdot\sigma(t) + \frac{\partial\left(\sqrt{\gamma^2(y,z)-\sigma^2(t)}\right)}{\partial y} \right\} \qquad (13)$$

Taking into account the aforementioned formulae (10)-(13) as well, system of equations (7) should be reduced properly as below ($A = A(t)$):

$$\begin{cases} \dfrac{d}{dt}\left(\dfrac{\sqrt{\gamma^2(y,z)-\sigma^2(t)}}{\gamma(y,z)+\sin(Ax)\cdot\sigma(t)}\right) = \dfrac{1}{2}\left(-\dfrac{\partial\left(\sqrt{\gamma^2(y,z)-\sigma^2(t)}\right)}{\partial z}-A\cdot\cos(Ax)\cdot\sigma(t)\right)\cdot\left(\dfrac{\sqrt{\gamma^2(y,z)-\sigma^2(t)}}{\gamma(y,z)+\sin(Ax)\cdot\sigma(t)}\right)^2 - \\[4pt] \qquad -\dfrac{1}{2}\left(-\dfrac{\partial\left(\sqrt{\gamma^2(y,z)-\sigma^2(t)}\right)}{\partial z}-A\cdot\cos(Ax)\cdot\sigma(t)\right)\cdot\left(\dfrac{\cos(Ax)\cdot\sigma(t)}{(\gamma(y,z)+\sin(Ax)\cdot\sigma(t))}\right)^2 + \\[4pt] \qquad +\left(A\cdot\sin(Ax)\cdot\sigma(t)+\dfrac{\partial\left(\sqrt{\gamma^2(y,z)-\sigma^2(t)}\right)}{\partial y}\right)\cdot\left(\dfrac{\cos(Ax)\cdot\sigma(t)}{(\gamma(y,z)+\sin(Ax)\cdot\sigma(t))}\right)+\dfrac{1}{2}\left(-\dfrac{\partial\left(\sqrt{\gamma^2(y,z)-\sigma^2(t)}\right)}{\partial z}-A\cdot\cos(Ax)\cdot\sigma(t)\right), \\[10pt] \dfrac{d}{dt}\left(\dfrac{\cos(Ax)\cdot\sigma(t)}{(\gamma(y,z)+\sin(Ax)\cdot\sigma(t))}\right) = \left(-\dfrac{\partial\left(\sqrt{\gamma^2(y,z)-\sigma^2(t)}\right)}{\partial z}-A\cdot\cos(Ax)\cdot\sigma(t)\right)\cdot\left(\dfrac{\sqrt{\gamma^2(y,z)-\sigma^2(t)}}{\gamma(y,z)+\sin(Ax)\cdot\sigma(t)}\right)\cdot\left(\dfrac{\cos(Ax)\cdot\sigma(t)}{(\gamma(y,z)+\sin(Ax)\cdot\sigma(t))}\right) - \\[4pt] \qquad -\left(A\cdot\sin(Ax)\cdot\sigma(t)+\dfrac{\partial\left(\sqrt{\gamma^2(y,z)-\sigma^2(t)}\right)}{\partial y}\right)\cdot\left(\dfrac{\sqrt{\gamma^2(y,z)-\sigma^2(t)}}{\gamma(y,z)+\sin(Ax)\cdot\sigma(t)}\right). \end{cases} \quad (14)$$

Using presentation of solution (10)-(13), we should solve (14) as the system of *ordinary* differential equations in regard to the time-parameter *t* (for which variables {*x,y,z*} should be considered as variable parameters).

Obviously, system of equations (14) is the system of *two non-linear* ordinary differential equations of the 1-st order for two functions *A(t)*, *σ(t)* (which could be solved by numerical methods only).

But nevertheless, the solution of system (14) exists, that's why solution of the system of equations (3) should exist also (for the presented, previously chosen form of the solution (5)+(6)). Approximate solutions of (14) are presented in the next section.

### 4. **Final presentation of the solution.**

Let us present the non-stationary exact solution {*p*, ***u***} of the Euler equations (3) in its final form, which should conserve the *Bernoulli*-function (*B = const*) (5) to be invariant for the aforementioned system:

$$\frac{\nabla p}{\rho} = -\nabla \phi - \frac{1}{2}\nabla\{(\vec{u})^2\}, \quad \{\nabla \cdot \vec{u} = 0\}, \quad \vec{u} \equiv \{u_1, u_2, u_3\}, \tag{15}$$

$$u_1 = -\sqrt{\gamma^2(y,z) - \sigma^2(t)}, \quad u_2 = -\lambda(x,t), \quad u_3 = \sigma(x,t),$$

$$\gamma \equiv \gamma(y,z), \quad \sigma(x,t) \equiv \sin(A(t) \cdot x) \cdot \sigma(t), \quad \lambda(x,t) \equiv \cos(A(t) \cdot x) \cdot \sigma(t),$$

- where $\rho$ is the fluid density, $\phi$ is the potential of external force, acting on a fluid; function $\gamma(y, z)$ is some arbitrary function, given by the initial conditions, $\gamma^2 > \sigma^2(t)$; *the time-dependent* functions $A(t)$, $\sigma(t)$ in the components of solution (15) are supposed to be the appropriate solutions of system of *two non-linear* ordinary differential equations of the 1-st order (14) (variables $\{x,y,z\}$ should be considered as variable parameters for them; components of the curl field in (14) are given properly by the expressions (13)).

Let us especially note that function $\gamma(x,y,z)$ along with the functions $\{u_1, u_2, u_3\}$ are proved to satisfy to the continuity equation (1) for such an expressions.

System of equations (14) could be transformed as below (let us note again that we consider variables $\{x,y,z\}$ as variable parameters):

$$\begin{cases}
\dfrac{-\dfrac{2\sigma(t)\,\sigma'}{2\sqrt{\gamma^2(y,z)-\sigma^2(t)}}\cdot(\gamma(y,z)+\sin(Ax)\cdot\sigma(t)) \;-\; (A'x\cdot\cos(Ax)\cdot\sigma(t)+\sin(Ax)\cdot\sigma')\cdot\sqrt{\gamma^2(y,z)-\sigma^2(t)}}{(\gamma(y,z)+\sin(Ax)\cdot\sigma(t))^2} = \\[2mm]
= \dfrac{1}{2}\left(-\dfrac{\partial\left(\sqrt{\gamma^2(y,z)-\sigma^2(t)}\right)}{\partial z}-A\cdot\cos(Ax)\cdot\sigma(t)\right)\cdot\left(\dfrac{\sqrt{\gamma^2(y,z)-\sigma^2(t)}}{\gamma(y,z)+\sin(Ax)\cdot\sigma(t)}\right)^2 - \\[2mm]
\quad -\dfrac{1}{2}\left(-\dfrac{\partial\left(\sqrt{\gamma^2(y,z)-\sigma^2(t)}\right)}{\partial z}-A\cdot\cos(Ax)\cdot\sigma(t)\right)\cdot\left(\dfrac{\cos(Ax)\cdot\sigma(t)}{(\gamma(y,z)+\sin(Ax)\cdot\sigma(t))}\right)^2 + \\[2mm]
\quad +\left(A\cdot\sin(Ax)\cdot\sigma(t)+\dfrac{\partial\left(\sqrt{\gamma^2(y,z)-\sigma^2(t)}\right)}{\partial y}\right)\cdot\left(\dfrac{\cos(Ax)\cdot\sigma(t)}{(\gamma(y,z)+\sin(Ax)\cdot\sigma(t))}\right) + \dfrac{1}{2}\left(-\dfrac{\partial\left(\sqrt{\gamma^2(y,z)-\sigma^2(t)}\right)}{\partial z}-A\cdot\cos(Ax)\cdot\sigma(t)\right), \\[4mm]
\dfrac{(-A'x\cdot\sin(Ax)\cdot\sigma(t)+\cos(Ax)\cdot\sigma')\cdot(\gamma(y,z)+\sin(Ax)\cdot\sigma(t)) \;-\; (A'x\cdot\cos(Ax)\cdot\sigma(t)+\sin(Ax)\cdot\sigma')\cdot\cos(Ax)\cdot\sigma(t)}{(\gamma(y,z)+\sin(Ax)\cdot\sigma(t))^2} = \\[2mm]
= \left(-\dfrac{\partial\left(\sqrt{\gamma^2(y,z)-\sigma^2(t)}\right)}{\partial z}-A\cdot\cos(Ax)\cdot\sigma(t)\right)\cdot\left(\dfrac{\sqrt{\gamma^2(y,z)-\sigma^2(t)}}{\gamma(y,z)+\sin(Ax)\cdot\sigma(t)}\right)\cdot\left(\dfrac{\cos(Ax)\cdot\sigma(t)}{(\gamma(y,z)+\sin(Ax)\cdot\sigma(t))}\right) - \\[2mm]
\quad -\left(A\cdot\sin(Ax)\cdot\sigma(t)+\dfrac{\partial\left(\sqrt{\gamma^2(y,z)-\sigma^2(t)}\right)}{\partial y}\right)\cdot\left(\dfrac{\sqrt{\gamma^2(y,z)-\sigma^2(t)}}{\gamma(y,z)+\sin(Ax)\cdot\sigma(t)}\right),
\end{cases}$$

(16)

- or, in other form

$$\begin{cases}
-\sigma' \cdot \gamma(y,z) \cdot \Big( \sigma(t) + \sin(Ax) \cdot \gamma(y,z) \Big) - A'x \cdot \cos(Ax) \cdot \sigma(t) \cdot \Big( \gamma^2(y,z) - \sigma^2(t) \Big) = \\[6pt]
= -\dfrac{1}{2}\sqrt{\gamma^2(y,z)-\sigma^2(t)} \cdot \left( \dfrac{\partial\left(\sqrt{\gamma^2(y,z)-\sigma^2(t)}\right)}{\partial z} + A \cdot \cos(Ax)\cdot \sigma(t) \right) \cdot \Big( \gamma^2(y,z) - \sigma^2(t) - \cos^2(Ax)\cdot \sigma^2(t) \Big) + \\[6pt]
+ \sqrt{\gamma^2(y,z)-\sigma^2(t)} \cdot \left( A\cdot \sin(Ax)\cdot \sigma(t) + \dfrac{\partial\left(\sqrt{\gamma^2(y,z)-\sigma^2(t)}\right)}{\partial y}\right) \cdot \cos(Ax)\cdot \sigma(t)\cdot(\gamma(y,z)+\sin(Ax)\cdot\sigma(t)) + \\[6pt]
+ \dfrac{1}{2}\sqrt{\gamma^2(y,z)-\sigma^2(t)} \cdot \left( -\dfrac{\partial\left(\sqrt{\gamma^2(y,z)-\sigma^2(t)}\right)}{\partial z} - A\cdot \cos(Ax)\cdot\sigma(t)\right)\cdot (\gamma(y,z)+\sin(Ax)\cdot\sigma(t))^2 ,
\end{cases}$$

(17)

$$\begin{cases}
-A'x\cdot\sigma(t)\cdot\Big(\sigma(t)+\sin(Ax)\cdot\gamma(y,z)\Big) + \cos(Ax)\cdot\gamma(y,z)\cdot\sigma' = \\[6pt]
= \left(-\dfrac{\partial\left(\sqrt{\gamma^2(y,z)-\sigma^2(t)}\right)}{\partial z} - A\cdot\cos(Ax)\cdot\sigma(t)\right)\cdot\sqrt{\gamma^2(y,z)-\sigma^2(t)}\cdot\cos(Ax)\cdot\sigma(t) - \\[6pt]
- \left(A\cdot\sin(Ax)\cdot\sigma(t) + \dfrac{\partial\left(\sqrt{\gamma^2(y,z)-\sigma^2(t)}\right)}{\partial y}\right)\cdot\sqrt{\gamma^2(y,z)-\sigma^2(t)}\cdot(\gamma(y,z)+\sin(Ax)\cdot\sigma(t)).
\end{cases}$$

But if we take into account the set of additional assumptions $A \to 0$, $\sigma \to 0$ in (17) (besides, $A(t)$ is supposed to be the slowly varying function depending on *t*), the aforementioned assumptions should reduce (16)-(17) as below:

$$\begin{cases}
-\sigma'\cdot\gamma(y,z)\cdot\sigma(t) - \sigma'\cdot\gamma^2(y,z)\cdot Ax - A'x\cdot\sigma(t)\cdot\gamma^2(y,z) + A'x\cdot\sigma^3(t) = \\[6pt]
= -\dfrac{\partial\gamma(y,z)}{\partial z}\cdot\gamma^3(y,z) + \dfrac{\partial\gamma(y,z)}{\partial y}\cdot\sigma(t)\cdot\gamma^2(y,z) , \\[12pt]
-A'x\cdot\sigma^2(t) - A'x\cdot\sigma(t)\cdot Ax\cdot\gamma(y,z) + \gamma(y,z)\cdot\sigma' = \\[6pt]
= -\gamma(y,z)\cdot\dfrac{\partial\gamma(y,z)}{\partial z}\cdot\sigma(t) - \dfrac{\partial\gamma(y,z)}{\partial y}\cdot\gamma^2(y,z) ,
\end{cases}$$

(18)

- or, it can be transformed properly

$$\begin{cases} \dfrac{dA}{dt} = \dfrac{1}{\sigma(t)} \cdot \left( \dfrac{\partial \gamma(y,z)}{\partial y} \cdot \gamma(y,z) \cdot A + \dfrac{\partial \gamma(y,z)}{\partial z} \cdot \dfrac{\gamma(y,z)}{x} \right) , \\ \\ \dfrac{d\sigma}{dt} = -\dfrac{\partial \gamma(y,z)}{\partial z} \cdot \sigma(t) - \dfrac{\partial \gamma(y,z)}{\partial y} \cdot \gamma(y,z) , \end{cases} \quad (19)$$

$\Rightarrow$

$$\begin{cases} A(t) = \dfrac{\exp\left( D \cdot \gamma(y,z) \cdot \int \left(\dfrac{1}{\sigma(t)}\right) dt \right) - B \cdot \dfrac{\gamma(y,z)}{x}}{D \cdot \gamma(y,z)} , \\ \\ \sigma(t) = -\left( \dfrac{\exp(-B \cdot (t-t_0)) + D \cdot \gamma(y,z)}{B} \right) \end{cases} \quad (20)$$

- here we denote $\partial\gamma/\partial z = B$, $\partial\gamma/\partial y = D$; so, we have presented in (20) the approximate solutions of system (14), where expression

$$A(t) = \dfrac{\exp(-B \cdot t_0)}{D \cdot \gamma(y,z)} \cdot \left( \dfrac{\exp(-B \cdot t)}{(D \cdot \gamma(y,z) \cdot \exp(-B \cdot t_0) + \exp(-B \cdot t))} - \dfrac{\gamma(y,z)}{x} \cdot B \right)$$

- means that $A(t)$ is proved to be the slowly varying function depending on $t$, indeed (according to the previously suggested assumption); variables $\{x,y,z\}$ should be considered as variable parameters in (20).

Let us schematically imagine at Figs. 1-3 the appropriate components of velocity field {$u_1$, $u_2$, $u_3$}, according to the formulae (15), which correspond to the approximate solutions (20) for functions $A(t)$, $\sigma(t)$ via derivation (18)-(19).

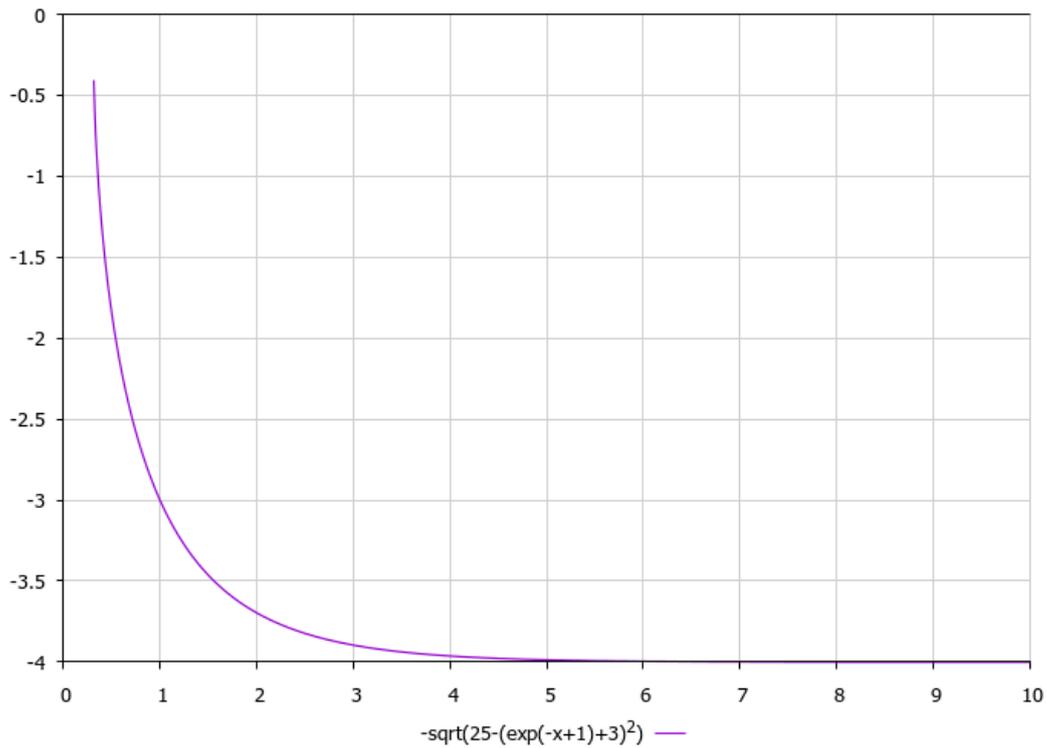

Fig.1. A *schematic* plot of the component $u_1$ (t)

of velocity field (15), depending on time $t$ (for simplicity, $B = 1$)

$$u_1 = -\sqrt{\gamma^2(y,z) - \left(\exp\left(-(t-t_0)\right) + D\cdot\gamma(y,z)\right)^2}$$

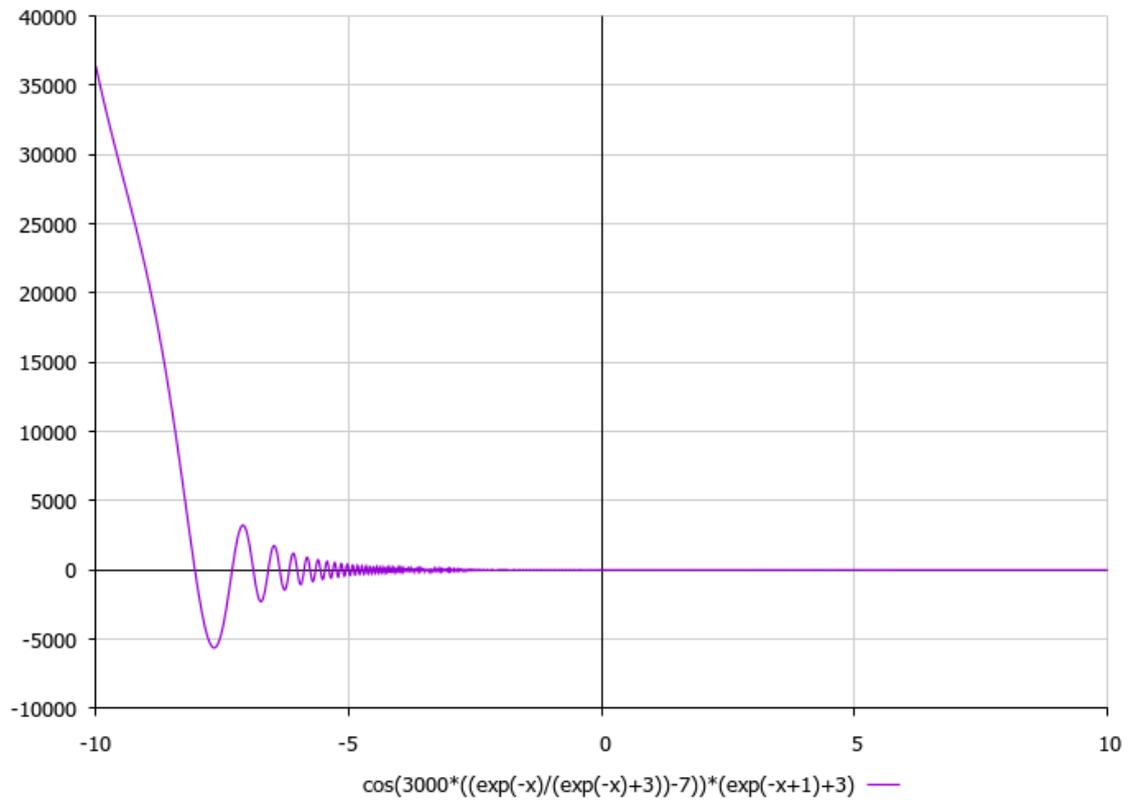

Fig.2. A *schematic* plot of the component $u_2(t)$

of velocity field (15), depending on time $t$ (for simplicity, $B = 1$)

$$u_2 = -\cos(A(t)\cdot x)\cdot \sigma(t), \quad \Rightarrow$$

$$\Rightarrow \quad u_2 = \cos(A(t)\cdot x)\cdot \left(\exp\left(-(t-t_0)\right) + D\cdot \gamma(y,z)\right),$$

$$A(t) = \frac{\exp(-t_0)}{D\cdot \gamma(y,z)}\cdot \left(\frac{\exp(-t)}{\left(D\cdot \gamma(y,z)\cdot \exp(-t_0) + \exp(-t)\right)} - \frac{\gamma(y,z)}{x}\right)$$

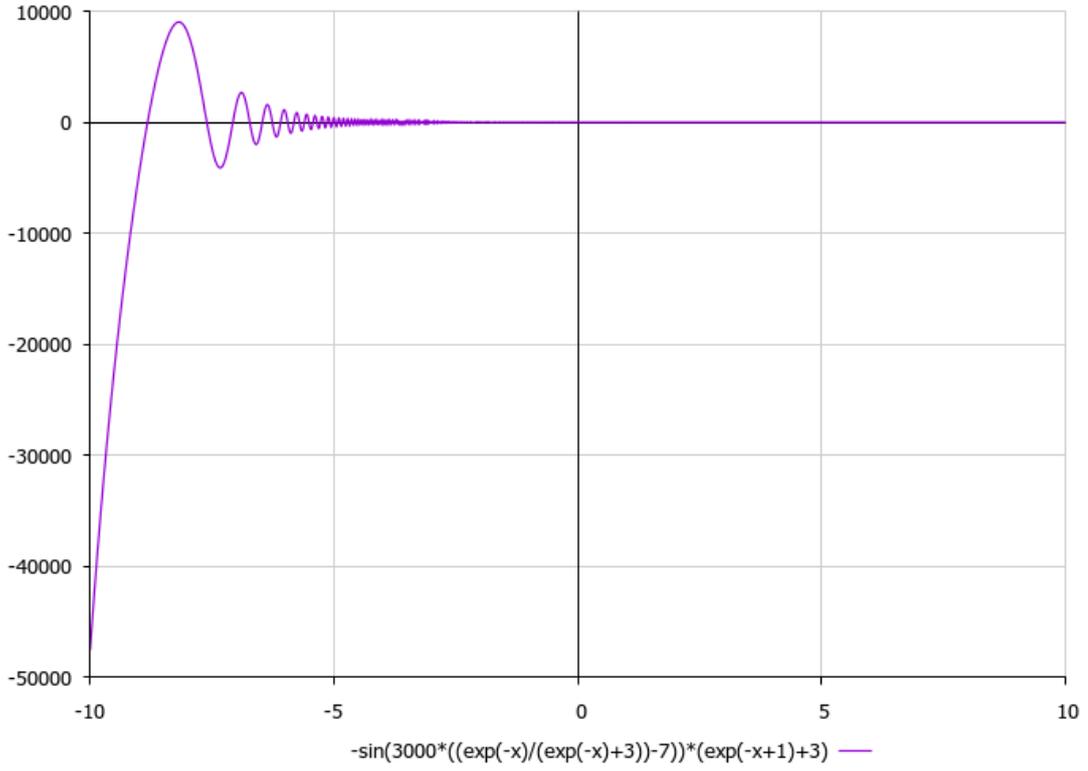

Fig.3. A *schematic* plot of the component $u_3$ (*t*)

of velocity field (15), depending on time *t* (for simplicity, *B = 1*)

$$u_3 \equiv \sigma(x,t) = \sin(A(t)\cdot x)\cdot \sigma(t), \Rightarrow$$

$$\Rightarrow u_3 = -\sin(A(t)\cdot x)\cdot\left(\exp(-(t-t_0)) + D\cdot\gamma(y,z)\right),$$

$$A(t) = \frac{\exp(-t_0)}{D\cdot\gamma(y,z)}\cdot\left(\frac{\exp(-t)}{(D\cdot\gamma(y,z)\cdot\exp(-t_0)+\exp(-t))} - \frac{\gamma(y,z)}{x}\right)$$

# 5. Discussion and conclusion.

Meanwhile, the intriguing fact is that Riccati-type seems to be the specific feature for a lot of the solutions in fluid mechanics and magneto-hydrodynamics [11]-[12]. We should mention very interesting results of article [13], where author discusses the presentation of components of the time-dependent solution for 3D Euler equation, which are supposed to be associated with complex Riccati equation via quaternionic formulation.

We should also mention the paper [14], in which author provides an example of the Riccati equation without periodic solutions which appears in the Euler vorticity dynamics.

All in all, Euler and Navier-Stokes equations have already been investigated in many researches including their numerical and analytical solutions [15]. However essential deficiency exists in the studies of non-stationary solutions of even the Euler equations for ideal, inviscid flow.

In this derivation, we explore the case of non-stationary flows of the Euler equations for incompressible fluids, which should conserve the *Bernoulli*-function for the flow.

The aforementioned approach quite differs from that which was used previously in investigation of Euler equations [16] by one of the authors. Indeed, an exuberant assumption was made in [16], such as separating of variables for the presenting of each components of the solution (including the pressure field). As for the relevance of such the purely mathematical assumption, it seems not to be physically reasonable if we will compare it to the approach for obtaining of our new solution with constant *Bernoulli*-function which should be invariant inside the limited domain of the flow.

To obtain new solution, we use previously suggested ansatz for solving of the system of Navier-Stokes equations (which is proved to have *the analytical* way to present its solution in case of conserving the *Bernoulli*-function for this type of the flows). Conditions for the existence of exact solution of the aforementioned type are proposed. The restrictions at choosing of the form of 3D solution for the given constant *Bernoulli*-function $B$ are emphasized: for example, the pressure field should be calculated from the given constant *Bernoulli*-function *B*, if all the components of velocity field are obtained.

Besides, we should especially note that in our case of solution, we need some boundary conditions that preserve the aforementioned solution inside of the chosen domain: indeed, the pressure field should strongly depend on the spatial part of the initial conditions for the components of velocity field (according to the *Bernoulli* invariant).

Also we should note that since the fluid is incompressible for the development above, there is a strong link between boundary conditions and the solution inside.

The uniqueness of the presented solutions is not considered. In this respect we confine ourselves to mention the paper [17], in which all the difficulties concerning the uniqueness in unbounded domain are remarked (which should be the same for analytical solutions in inviscid fluids if the kinematic viscosity tends to zero).

The last but not least, we have found an elegant way to simplify the presented solution up to the analytical presentation of the approximate solution. Also it has to be specified that the solutions that are constructed can be considered as a class of perturbations, absorbed exponentially as *t* going to infinity ∞ by the null solution.

## **Acknowledgements**

Authors are thankful to Dr. A.V.Koptev [18], to Dr. V.I.Semenov [19], and especially to Dr. G.V.Alekseev [20] with respect to useful discussions during preparing of the PhD-thesis (of one of the authors) as well as for their wise advices regarding the further works (including this manuscript).

We should also appreciate the efforts of 3 esteemed reviewers for their valuable advices which improved structure of the article significantly.

This study was initiated in the framework of the state task programme in the sphere of scientific activity of the Ministry of Education and Science of the Russian Federation (project No. 5.5176.2017/8.9) and grant of the President of the Russian Federation for state support of the leading scientific schools of the Russian Federation (NSh-6637.2016.5).

**References:**

[1]. Landau, L.D.; Lifshitz, E.M. (1987), *Fluid mechanics, Course of Theoretical Physics 6* (2nd revised ed.), Pergamon Press, ISBN 0-08-033932-8.

[2]. Ladyzhenskaya, O.A. (1969), *The Mathematical Theory of viscous Incompressible Flow* (2nd ed.), Gordon and Breach, New York.

[3]. Lighthill, M. J. (1986), *An Informal Introduction to Theoretical Fluid Mechanics*, Oxford University Press, ISBN 0-19-853630-5.

[4]. Saffman, P. G. (1995), *Vortex Dynamics*, Cambridge University Press.

[5]. Milne-Thomson, L.M. (1950), *Theoretical hydrodynamics*, Macmillan.

[6]. Kamke E. (1971), *Hand-book for Ordinary Differential Eq*. Moscow: Science.

[7]. Ershkov S.V. (2015). *On Existence of General Solution of the Navier-Stokes Equations for 3D Non-Stationary Incompressible Flow*, International Journal of Fluid Mechanics Research, 06/2015, 42(3), pp. 206-213. Begell House.

[8]. Ershkov S.V. (2015). *Non-stationary Riccati-type flows for incompressible 3D Navier-Stokes equations*, Computers and Mathematics with Applications, vol. 71, no. 7, pp. 1392–1404.

[9]. Ershkov S.V. (2016). *A procedure for the construction of non-stationary Riccati-type flows for incompressible 3D Navier-Stokes equations*, Rendiconti del Circolo Matematico di Palermo, Volume 65, Issue 1, pp.73-85.

[10]. Ershkov S.V. (2017). *A Riccati-type solution of Euler-Poisson equations of rigid body rotation over the fixed point*. Acta Mechanica, vol. 228, no. 7, pp. 2719–2723.

[11]. Christianto V., Smarandache F. (2008), *An Exact Mapping from Navier-Stokes Equation to Schroedinger Equation*, Progress in Physics, 1, p.38-39.

[12]. Gibbon, J.D., Holm D.D., Kerr R.M., and Roulstone I. (2006). *Quaternions and particle dynamics in the Euler fluid equations*, Nonlinearity, vol. 19 (8), pp. 1969–1983.

[13]. Gibbon, J.D. (2002). *A quaternionic structure in the three-dimensional Euler and ideal magneto-hydrodynamics equations*. Physica D, vol. 166, pp. 17–28.

[14]. Wilchynski, P. (2009). *Quaternionic-valued ordinary differential equations. The Riccati equation*. Journal of Differential Equations, vol. 247, pp. 2163–2187.

[15]. Drazin, P.G. and Riley N. (2006), *The Navier-Stokes Equations: A Classification of Flows and Exact Solutions*, Cambridge, Cambridge University Press.

[16]. Ershkov S.V. (2015). *Quasi-periodic non-stationary solutions of 3D Euler equations for incompressible flow*, Journal of King Saud University – Science, 06/2015.

[17]. Galdi G.P., Rionero S. (1979), *The weight function approach to uniqueness of viscous flows in unbounded domains*, Archive for Rational Mechanics and Analysis, Volume 69, Issue 1, pp 37-52.

[18]. Koptev A.V. (2014), *Generator of solutions for 2D Navier - Stokes equations*. Journal of Siberian Federal University. Mathematics & Physics. Issue 7(3). pp. 324-330.

[19]. Semenov V.I. (2014), *Some new identities for solenoidal fields and applications*. Mathematics. Issue 2. pp. 29-36; doi:3390/math2010029

[20]. Alekseev G.V. (2016), *Solvability of an inhomogeneous boundary value problem for the stationary magnetohydrodynamic equations for a viscous incompressible fluid*, Differential Equations, Vol.52, Issue 6, pp. 739-748.